\documentclass[twoside,9pt]{article}
\usepackage{extsizes}
\usepackage[super,sort&compress,comma]{natbib} 
\usepackage[version=3]{mhchem}
\usepackage[left=2.5cm, right=2.5cm, top=2cm, bottom=2.0cm]{geometry}
\usepackage{balance}
\usepackage{times,mathptmx}
\usepackage{sectsty}
\usepackage{graphicx} 
\usepackage{lastpage}
\usepackage[format=plain,justification=justified,singlelinecheck=false,font={stretch=1.125,small,sf},labelfont=bf,labelsep=space]{caption}
\usepackage{float}
\usepackage{fancyhdr}
\usepackage{fnpos}
\usepackage[english]{babel}
\addto{\captionsenglish}{%
  \renewcommand{\refname}{Notes and references}
}
\usepackage{array}
\usepackage{droidsans}
\usepackage{charter}
\usepackage[T1]{fontenc}
\usepackage[usenames,dvipsnames]{xcolor}
\usepackage{setspace}
\usepackage[compact]{titlesec}
\usepackage{hyperref}
\usepackage{amssymb} 

\usepackage{multicol}

\usepackage{epstopdf}

\definecolor{cream}{RGB}{222,217,201}

\begin{document}

\pagestyle{fancy} 

\fancypagestyle{plain}{

}


\makeFNbottom
\makeatletter
\renewcommand\LARGE{\@setfontsize\LARGE{15pt}{17}}
\renewcommand\Large{\@setfontsize\Large{12pt}{14}}
\renewcommand\large{\@setfontsize\large{10pt}{12}}
\renewcommand\footnotesize{\@setfontsize\footnotesize{7pt}{10}}
\makeatother

\renewcommand{\thefootnote}{\fnsymbol{footnote}}
\renewcommand\footnoterule{\vspace*{1pt}%
\color{cream}\hrule width 3.5in height 0.4pt \color{black}\vspace*{5pt}} 
\setcounter{secnumdepth}{5}

\makeatletter 
\renewcommand\@biblabel[1]{#1}            
\renewcommand\@makefntext[1]%
{\noindent\makebox[0pt][r]{\@thefnmark\,}#1}
\makeatother 
\renewcommand{\figurename}{\small{Fig.}~}
\sectionfont{\sffamily\Large}
\subsectionfont{\normalsize}
\subsubsectionfont{\bf}
\setstretch{1.125} 
\setlength{\skip\footins}{0.8cm}
\setlength{\footnotesep}{0.25cm}
\setlength{\jot}{10pt}
\titlespacing*{\section}{0pt}{20pt}{4pt}
\titlespacing*{\subsection}{0pt}{15pt}{1pt}

\fancyfoot{}
\fancyfoot[RO]{\footnotesize{\sffamily{1--\pageref{LastPage} ~\textbar  \hspace{2pt}\thepage}}}
\fancyfoot[LE]{\footnotesize{\sffamily{\thepage~\textbar\hspace{3.45cm} 1--\pageref{LastPage}}}}
\fancyhead{}
\renewcommand{\headrulewidth}{0pt} 
\renewcommand{\footrulewidth}{0pt}
\setlength{\arrayrulewidth}{1pt}
\setlength{\columnsep}{6.5mm}
\setlength\bibsep{1pt}

\makeatletter 
\newlength{\figrulesep} 
\setlength{\figrulesep}{0.5\textfloatsep} 

\newcommand{\topfigrule}{\vspace*{-1pt}%
\noindent{\color{cream}\rule[-\figrulesep]{\columnwidth}{1.5pt}} }

\newcommand{\botfigrule}{\vspace*{-2pt}%
\noindent{\color{cream}\rule[\figrulesep]{\columnwidth}{1.5pt}} }

\newcommand{\dblfigrule}{\vspace*{-1pt}%
\noindent{\color{cream}\rule[-\figrulesep]{\textwidth}{1.5pt}} }

\makeatother

\bigbreak

\begin{center}

\LARGE{Destabilization and phase separation of particle suspensions in emulsions}

\medbreak

\large{Blandine Feneuil,$^{\ast}$\textit{$^{a}$} Atle Jensen\textit{$^{a}$} and Andreas Carlson\textit{$^{a}$}}
\end{center}

\bigbreak

\subsection*{Abstract}
\noindent\normalsize{Yield stress fluids are widely used in industrial application to arrest dense solid particles, which can be studied by using a concentrated emulsion as a model fluid. We show in experiments that particle sedimentation in emulsions cannot be predicted by the classical criterion for spheres embedded in a yield stress fluid. Phase separation processes take place, where a liquid layer forms and particle sedimentation is enhanced by the emulsion drainage. In addition, emulsion drainage can be arrested or enhanced by the amount of particles embedded in the emulsion. A minimal mathematical model is developed and solved in numerical simulations to describe the emulsion drainage in the presence of particles, which favorably compares with the experimental stability diagram and the sedimentation dynamics..} \\


\renewcommand*\rmdefault{bch}\normalfont\upshape
\rmfamily
\section*{}
\vspace{-1cm}


\footnotetext{\textit{$^{a}$~ University of Oslo, Oslo, Norway.\\ E-mails: bffeneui@math.uio.no, atlej@math.uio.no, acarlson@math.uio.no}}


\section{Introduction}

Yield stress fluids only flow like liquids when the applied stress is above the yield stress, and they exhibit a solid behavior otherwise. This property offers many advantages which are relevant to industrial applications: toothpaste remains at the top of the toothbrush fibers \cite{2018_Ahuja}, paint can stick to vertical walls or fibers \cite{2019_Smit} and bread dough can be shaped at will by the baker \cite{2008_Sofou}. One of these many advantages is the ability to avoid sedimentation of dense particles in the fluid at rest, crucial to avoid segregation of cement, sand and gravel in concrete \cite{2017_Massoussi}, or to ensure uniform distribution of barite particles in drilling fluids \cite{2002_Saasen}. 

In 1985, \textit{Beris et al.} \cite{1985_Beris} established theoretically a criterion for the sedimentation of a single rough spherical particle in an infinite yield stress fluid: the particle remains at rest if the suspending fluid's yield stress $\tau_y$ exceeds a critical value $\tau_{y,s}$ given by the following formula:
\begin{equation}
\tau_{y,s} = \dfrac{1}{21} (\rho_P - \rho_f) g d,
\label{equation_criterion}
\end{equation}
where $\rho_P$ and $\rho_f$  are the density of the particle and the suspending yield stress fluid respectively, and $d$ is the diameter of the particle. This criterion was later verified experimentally \cite{2007_Tabuteau}. Extentions to more complex configurations have been studied experimentally and numerically: the presence of a vertical wall in the vicinity of the particle does not affect the criterion of Eq. \ref{equation_criterion} \cite{1997_Blackery, 2004_Mitsoulis}. However, $\tau_{y,s}$ is affected by the shape and orientation of the particle: a fiber or spheroid sediments more easily (i.e. the critical yield stress is higher) if the elongation direction is vertical than if it is horizontal \cite{2001_Jossic, 2010_Putz}. In addition, for smooth particles, i.e. when wall slip can occur on the solid surface, $\tau_{y,s}$  increases \cite{2001_Jossic,2003_Deglo,2006_Merkak}. Another crucial step toward understanding particle sedimentation in complex industrial materials is the presence of several particles. For two spheres aligned horizontally, the criterion is not affected. If the spheres are aligned vertically, they can interact, which enhances sedimentation \cite{2003_Liu, 2006_Jie, 2009_Tokpavi_1}, requiring a larger yield stress to prevent sedimentation \cite{2006_Merkak,2009_Jossic}. $\tau_{y,s}$  depends on the interparticle distance: for separation distance larger than 4 times the particle diameter, Eq. \ref{equation_criterion} is still valid, and the value increases when the distance is decreasing and is maximal when spheres are in contact \cite{2018_Chaparian}. The maximal value is less than twice the critical value from Eq. \ref{equation_criterion}. When three or more particles are present, a new phenomena can be observed: some particles can be trapped while others are not, leading to particle agglomeration \cite{2018_Chaparian}. The number of trapped particles depends on the yield stress and on the particle distance.

\begin{figure*}
\centering
\includegraphics[width=0.9\textwidth]{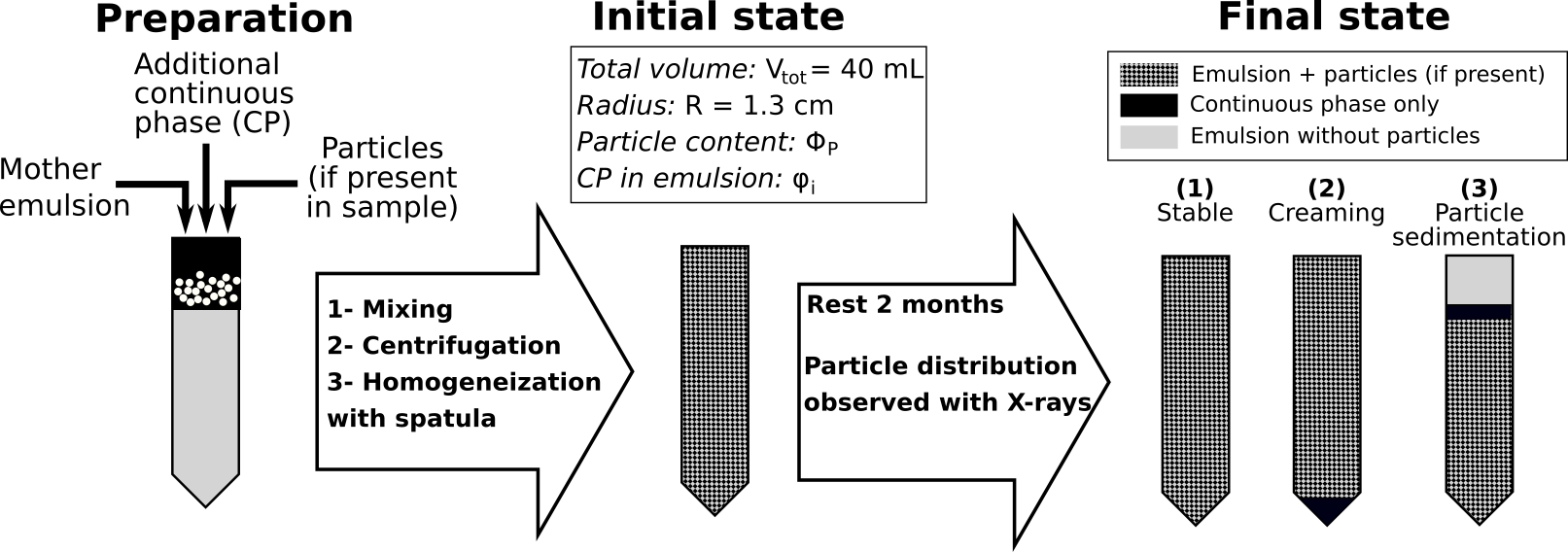}
\caption{A schematic description of the experimental methodology. 40-mL samples are prepared by mixing the ``mother" emulsion, some additional continuous phase and glass particles, with proportions chosen to obtain uniform values of particle content $\Phi_p$ and continuous phase content in emulsion $\varphi_i$. To track the particle distribution in space and time we perform X-ray experiments, and the stability of the mixture is evaluated after two months at rest.}
\label{schema_method}
\end{figure*}

Concentrated emulsions, i.e. a dispersion of Newtonian liquid droplets embedded in another immiscible Newtonian liquid, offer one way to create a material with yield stress properties \cite{1983_Princen,1989_Princen}. When looking at flow phenomena at length scales much greater than the droplet size, these emulsions can be considered as simple yield stress fluids \cite{2013_Ovarlez}: no shear localisation occurs, and the yield stress does not depend on shear history. ``Concentrated" means that the volume fraction of the droplets is so large that they are not spherically shaped. The yield stress arises from the fact that, to make the material flow, an applied stress must be large enough to allow each droplet to escape from the "cage" formed by its neighbors \cite{1983_Princen}. Yield stress behavior therefore appears when the volume fraction of the continuous phase, $\varphi$, is below a critical value $\varphi_c$. The critical volume fraction $\varphi_c$ depends on the droplet size distribution \cite{2009_Farr}. To quantify the deformation of droplets in concentrated emulsions, it is convenient to use the notion of osmotic pressure, $\Pi$, which was introduced by \textit{Princen} in 1979 \cite{1979_Princen}. It is the pressure that must be applied on the emulsion droplets to reach a given $\varphi$ when the continuous phase can freely move between the emulsion and a reservoir. Foams, i.e. dispersion of gas bubbles in a liquid, have very similar features as emulsions and many of the same principles apply \cite{2013_Ovarlez,2013_Maestro}.

Earlier work with emulsions and foams have pointed out three mechanisms than can destabilize the material \cite{2013_Tadros,2013_Cantat}. Droplet (or bubble) coalescence and Ostwald ripening, i.e. transfer of molecules from the smaller droplets to the bigger, both lead to an increase of the average droplet size. The third mechanism, drainage, is due to gravitational effects. If the droplets are lighter than the continuous phase, they rise in the emulsion. In this scenario, $\varphi$ will increase at the bottom and decrease at the top of the emulsion. In a vertical column of emulsion at equilibrium, for each plane at a given height z, pressure due to the buoyancy of the droplets below the plane must be in equilibrium with the osmotic pressure of the emulsion above. The osmotic pressure is therefore related to the volume fraction of the continuous phase \cite{1986_Princen,2013_Maestro}.

Sedimentation of particles in foams or emulsions depends on the relative droplet and particle size. Very small particles can be transported in channels and nodes between the dispersed phase \cite{2010_Rouyer_b}. When the particle size increases, they can be retained in the nodes \cite{2014_Khidas}. When particles are very large compared to the dispersed phase, the emulsion behaves like a continuous material at the particle scale, and we would expect that criterion of Eq. \ref{equation_criterion} applies, which we examine here. We show experimentally that the Eq. \ref{equation_criterion} cannot predict sedimentation of particles if emulsions are considered as homogeneous fluids. Emulsion drainage favors particle sedimentation, and the presence of particles can either prevent or enhance drainage. This behavior can be captured by a minimal model, with takes into account the coupling between drainage and particle sedimentation. The model is used to predict the final stability of the samples, and to study the destabilization kinetics.

\section{Materials and methods}

The experimental procedure is summarized in Fig. \ref{schema_method}. All samples are prepared from a single ``mother" emulsion, which preparation is described in the next paragraph. After particles are added and the sample is homogeneized, it is left at rest for two months at least. During the resting time, we observe samples with X-rays to measure particle sedimentation and phase-separation of the emulsion.

\subsection{``Mother" emulsion}

The ``mother" oil-in-water emulsion is prepared as follows. First, 3wt\% TTAB surfactant (tetradecyltrymethylammonium bromide, from VWR) is dissolved in the continuous phase containing 30wt\% of glycerol in deionized water, with density $\rho_{CP}=$1065~kg/m$^3$ and viscosity $\mu =$ 2.46 mPa.s. A very small amount of Rhodamin dye is added to the continuous phase for some samples, giving it a pink color, allowing us to identify phases when phase-separation occurs.
Then silicon oil (V350, from VWR), with a density $\rho_{oil} = 970 $ kg/m$^3$, is added and droplets are obtained with a Silverson emulsifier, by increasing the rotation velocity of the mixer from 600 rpm to 2400 rpm by steps of 600 rpm. The volume fraction of the continuous phase in the mother emulsion is 16.5\%. 
The interfacial tension $\gamma$ of oil-water interfaces stabilized with TTAB is $\gamma \approx 7.5$ mN/m\cite{2009_Georgieva}. 
We have confirmed the reproducibility of our preparation protocol by measuring the rheological properties of two emulsions prepared independently.

\begin{figure}[h]
\centering
\includegraphics[width=6.6cm]{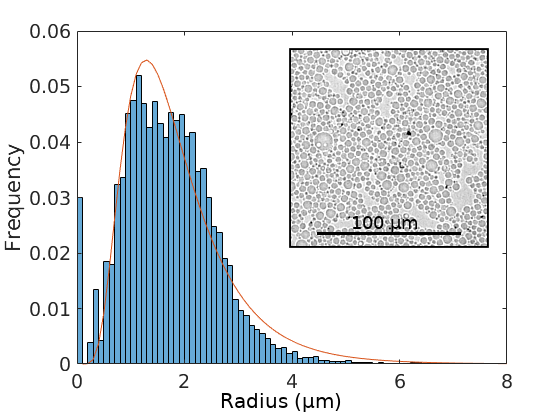}
\caption{Size distribution of droplets in the emulsion. The blue histogram has been obtained for about 25000 droplets and the red curve is a log-normal fit $ P(R) = exp(-[ln(R/R_0)]^2/2\sigma^2)/\sqrt{2\pi}R$, with average radius $R_0 = 1.65~\mu$m and a standard deviation $\sigma = 0.5$. Inset: A microscope image of the diluted emulsion.}
\label{histogram_emulsion}
\end{figure}

The droplet size distribution has been measured by analysis of microscope images (see Fig. \ref{histogram_emulsion}). The emulsion is diluted by adding some continuous phase to facilitate the visualization of the droplets. From these images we have also calculated the mean Sauter radius of the droplets: $R_{32} = <R^3>/<R^2> = 2.6~\mu$m, where $<R^3>$ and $<R^2>$ are the averages of respectively the cube and the square of the droplet radii $R$.

\subsection{Particle sedimentation in stagnant condition}

To obtain the 40~mL samples out of the mother emulsion, we add continuous phase to dilute the emulsion to the desired oil volume fraction, and glass particles. 
Glass particle are mixed with the additional continuous phase prior to incorporation into the mother emulsion to avoid entrainment of air bubbles. We use spherical glass particles with diameters d = [100, 400] $\mu$m and of density $\rho_P =$ 2500 kg/m$^3$. Before the experiments, they are washed with aceton to remove impurities, and then with isopropanol to remove traces of oil. Note that the diameter of the glass spheres is two order of magnitude larger than the diameter of the emulsion droplets, so that the emulsion is a continuous material at the scale of the particles. Samples are mixed a first time, then centrifugated to remove the bubbles, and finally homogenized again with a spatula in a 50~mL tube. Then the tube is left at rest vertically during the length of the experiment, at least two months. Note that all experiments have been carried out at least twice, first without and then with centrifugation, and the same final stability results have been obtained, with the exception of two samples at the border between the stable and unstable regions. We have chosen to only present the results obtained with centrifugation. The tubes have a conical bottom (see Fig. \ref{schema_method}). Their diameter is 2.6 cm, it is large compared to the size of the particles, allowing us to neglect wall effects. $\Phi_P = V_{P}/(V_{CP}+V_{oil}+V_{P})$  is the volume fraction of particles in the samples and $\varphi_i = V_{CP}/(V_{CP}+V_{oil})$ is the initial volume fraction of the continuous phase in the emulsion, with $V_{CP}$ the volume of the continuous phase, $V_{oil}$ the volume of oil and $V_{P}$ the volume of particles.
We will see that when the samples are left at rest, the volume fraction of the continuous phase becomes inhomogeneous. We will denote by $\varphi$ the local volume fraction of the continuous phase in the emulsion.

\subsection{X-rays}

To visualize the sedimentation of particles, we deploy an X-ray measurement technique. The sample is placed vertically between an X-ray source and a detector. The association of the X-ray source with a collimator provides a conical beam with horizontal axis. X-rays are generated with an 8 mA intensity of X-rays, at either 38 kV or 46 kV. On pictures obtained with the low voltage (38 kV) we distinguish the particle-free part of the samples from the particle-laden parts. The higher voltage (46 kV) helps us to distinguish the local variations of the concentration of particles. The X-ray CMOS detector size is 1536*864 pixels, and on the images, the spatial resolution is 56 $\mu$m/px.

\subsection{Rheometry}
We have measured the yield stress of the emulsion without particles as a function of the volume fraction of the continuous phase. Measurement have been carried out using a 5-cm diameter striated plate-plate geometry with 1.5 mm gap in an Anton Paar MRC 702 MD Rheometer in single drive set up. After pre-shearing the emulsion at 50 s$^{-1}$ for 1 min, we apply a 0.01 s$^{-1}$ constant shear rate and plot the measured stress. Each start-of-flow curve presents a linear regime when the strain is below 30\%, where the stress is proportional to the strain. Then the curve reaches a plateau, where the value is the sample's yield stress.

\section{Results and discussion}

\subsection{Yield stress for the onset of sedimentation}

\begin{figure}[h]
\centering
\includegraphics[width=6.5cm]{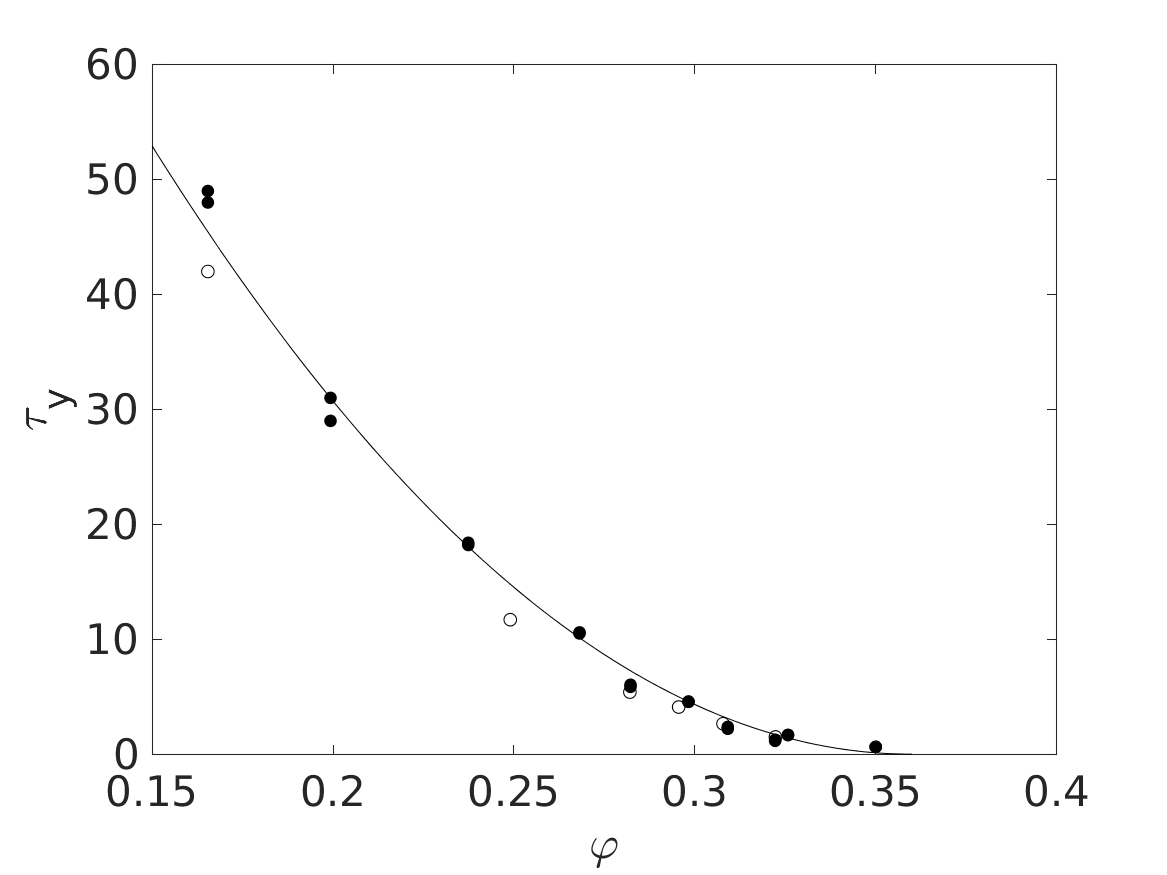}
\caption{The yield stress of emulsions as a function of the volume fraction of the continuous phase $\varphi$. Empty and filled circles refer to samples obtained from two different mother emulsions prepared following the same protocol. Black line is a fit with Eq. \ref{equation_YS_emulsion}: $\tau_y = \beta \dfrac{\gamma}{R_{32}}(\varphi-\varphi_c)^2$ with $\beta = 0.42$ and $\varphi_c = 0.36$. }
\label{graph_YS}
\end{figure}

The yield stress of the emulsion without particle is given as a function of the volume fraction of the continuous phase $\varphi$ in Fig. \ref{graph_YS}. We have checked the reproducibility of the preparation of the mother emulsion by measuring the rheological properties of samples obtained from two different mother emulsions, prepared separately using the same protocol. Values for both series of dilutions are on a single curve, which can be fitted by \cite{2013_Cantat}: 

\begin{equation}
\tau_y = \beta \dfrac{\gamma}{R_{32}}(\varphi_c-\varphi)^2,
\label{equation_YS_emulsion}
\end{equation}
where $\beta = 0.42$, and $\varphi_c=0.36$ is the minimal volume fraction of continuous phase where all the droplets would be spherical. This curve is in accordance with the literature on the yield stress of foams and emulsions, where $\varphi_c=0.36$ has been observed before, and values for the empirical constant $\beta$ are in the range 0.2-0.5 \cite{2013_Cantat}.

The yield stress threshold for particle sedimentation is calculated by Eq. \ref{equation_criterion}. Note that in our samples, as $\varphi$ varies from 0.24 to 0.34, the density of the emulsion varies only by 1\%, from 993 kg/m$^3$ to 1002 kg/m$^3$. Sedimentation of individual glass particles should not happen in emulsion if the yield stress is above 0.28 Pa when $d = 400 \mu$m, and 0.07 Pa when $d = 100 \mu$m. From Eq. \ref{equation_YS_emulsion}, we deduce the volume fraction $\varphi_s$ of continuous phase in emulsion above which sedimentation is expected: $\varphi_s$ = 0.345 for 400 $\mu$m particles and $\varphi_s$ = 0.352 for 100 $\mu$m particles. For all our samples, $\varphi_i < 0.34$; therefore, if the continuous phase remained uniformly distributed over the height of the samples, all the samples should be stable.

Our samples contain several particles. The number of particles in the sample is $N = \Phi_P V_{tot} /(\pi d^3/6)$. If they are uniformly distributed, sample volume per particle is $ V_{tot}/N = \pi d^3/(6 \Phi_P) \simeq \pi L^3/6$, where L is the interparticle distance. Thus, L can be estimated by $L/d \simeq 1/\sqrt[3]{\Phi_P}$. We obtain $ L/d \simeq 2 $ for $\Phi_P = 0.1$ and $ L/d \simeq 1.3 $ for $\Phi_P = 0.4$. This estimate shows that for most of our samples, the interparticle distance is of the same order as the particle diameter. We therefore need to keep in mind that a higher yield stress is required to arrest sedimentation that the value given by Eq. \ref{equation_criterion}. As far as we know, the critical value of yield stress is not known for a large number of particles. Values obtained for two particles where smaller that twice the value given by Eq. \ref{equation_criterion} \cite{2009_Jossic,2018_Chaparian}. We will discuss later the sensitivity of the model to the stability criterion.

\subsection{Model assumptions}

In the following, we will calculate how the volume fraction of the continuous phase changes during drainage. As long as at each height z, $\varphi < \varphi_s$, the sample is stable, i.e. the three phases (continuous phase, oil and particles) remain mixed. On the other hand, if $\varphi$ locally reaches high values, the sample starts to phase separate by two possible processes:

\begin{itemize}
\item $\varphi > \varphi_s$: particle sedimentation takes place and a layer of particle free emulsion appears.
\item $\varphi = \varphi_c$: a layer of continuous phase appears. Note that as $\varphi_s < \varphi_c$, if the layer of continuous phase appears, particle sedimentation also occurs.
\end{itemize}

In order to write a general model for both phase-separation phenomena, we note $\varphi_{dest}$ the critical volume fraction above which samples locally destabilize. $\varphi_{dest}$ can either refer to $\varphi_s$ or $\varphi_c$ depending on the phase separation process which is studied.

To calculate the evolution of the $\varphi$ profile due to emulsion drainage, we assume that particles are stuck by the droplets at each z position at long as $\varphi(z) < \varphi_{dest}$. The continuous phase can drain between the droplets and the particles. In other words, we study drainage in a material composed of a continuous phase of density $\rho_{CP}$ and a dispersed phase with an effective density $\rho_d$:

\begin{equation}
\rho_d = \dfrac{\rho_{oil} (1-\varphi)(1-\Phi_P) + \rho_P \Phi_P}{(1-\varphi)(1-\Phi_P) + \Phi_P}.
\label{equation_rho}
\end{equation}

\begin{figure}[h]
\centering
\includegraphics[width=5cm]{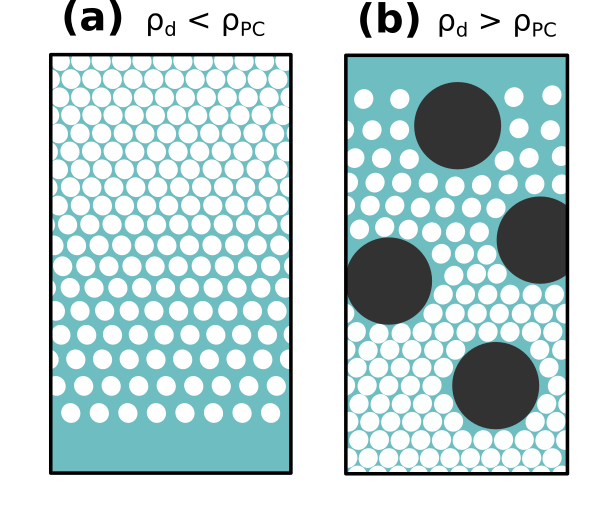}
\caption{Illustration of sample drainage. For (a), the density of the dispersed phase is smaller than the density of the continuous phase, and the continuous phase tends to flow toward the bottom of the sample. In (b), we illustrate, when $\rho_d > \rho_{PC}$: the continuous phase flows to the top of the sample. For the clarity of the drawing, the size of the droplets and the distance between them are not drawn to scale.} 
\label{schema_mechanism_static}
\end{figure}

The volume fraction of the continuous phase in this material is $\Phi = \varphi (1 - \Phi_P)$. All the macroscopic quantities used in this model, in particular $\Phi$ and the different pressure terms, are defined on a representative volume, with a size that is much larger than the particle size. From Eq. \ref{equation_rho}, we can calculate the $\Phi_P$ value at which dispersed and continuous phases have the same density. This value depends slightly on $\varphi$: if $\varphi_i$ = 0.24, we find $\Phi_P(\rho_d = \rho_{CP}) = 4.8\%$ and if $\varphi_i$ = 0.34, $\Phi_P(\rho_d = \rho_{CP}) = 4.2\%$. When $\rho_{CP} > \rho_d$  the drainage will lead to an increase of $\varphi$ at the bottom of the sample. On the contrary, when $\rho_{CP} < \rho_d$ the continuous phase flows towards the top of the sample. Both configurations are illustrated in Fig. \ref{schema_mechanism_static}.

\subsection{Static approach}

We start by calculating the equilibrium $\varphi$ profile by balancing the osmotic pressure and the hydrostatic pressure in the continuous phase. It has already been demonstrated that for a particle free emulsion, balance between gravity and the osmotic pressure leads to \cite{2013_Maestro,1986_Princen}:

\begin{equation}
\dfrac{d\Pi}{dz} = (1-\varphi) (\rho_{CP}-\rho_{d}) g  \text{\hspace{0.5 mm} when \hspace{0.3 mm}} \Phi_P = 0,
\label{equation_dOP}
\end{equation}
at each height z. By using our assumptions, we can generalize this expression to samples with particles as long as $\varphi < \varphi_s$:
\begin{equation}
\dfrac{d\Pi}{dz} = \left(1-\Phi\right) (\rho_{CP}-\rho_{d}) g = \left(1-\varphi(1-\Phi_P)\right) (\rho_{CP}-\rho_{d}) g.
\label{equation_dOP_2}
\end{equation}

Similarly to \textit{Princen} \cite{1986_Princen} and \textit{Maestro et al.} \cite{2013_Maestro}, we integrate Eq. \ref{equation_dOP_2} to find the relation between the dimensionless osmotic pressure $\tilde{\Pi}=\Pi / (\gamma/R_{32})$ and the dimensionless height $\tilde{z}=(R_{32}/\lambda_c^2)z$, with $\lambda_c = \sqrt{\gamma/|\rho_{CP}-\rho_d| g}$ the capillary length:

\begin{equation}
\tilde{z} = \int_0^{\tilde{\Pi}} \dfrac{d\tilde{\Pi}}{1-\varphi (1-\Phi_P)}.
\label{equation_integral_dOP}
\end{equation}

In this expression, we have defined the z axis so that $\tilde{z}=0$ when $\tilde{\Pi}=0$ (i.e. the droplets are spherical at z=0). In addition, $\varphi$ decreases when z increases, i.e. the z axis is oriented upwards when $\rho_d < \rho_{CP}$, and downwards when $\rho_d > \rho_{CP}$.

Next, we turn to an expression of the osmotic pressure which is arising from the increase of surface energy due to droplet deformation. Glass beads in our samples are non-deformable. In addition, their size is two orders of magnitude larger than the droplets, so they do not affect the droplet shape. We therefore assume that the osmotic pressure depends on $\varphi$ and is not related to $\Phi_P$. We will use the following empirical relation \cite{2008_Hohler,2013_Maestro}:

\begin{equation}
\tilde{\Pi} = k \dfrac{(\varphi_c-\varphi)^2}{\sqrt{\varphi}},
\label{equation_OP}
\end{equation}
where $k$ and $\varphi_c$ are empirical constants related by Princen's criterion \cite{1986_Princen,2013_Maestro}:

\begin{equation}
\int_0^{\varphi_c}\dfrac{\tilde{\Pi}(\varphi)}{(1-\varphi)^2}d\varphi = C.
\label{equation_criterion_Princen}
\end{equation}
The constant C accounts for the increase of area of the droplets when $\varphi\rightarrow 0$ and is further discussed in \cite{2013_Maestro}. For disordered dispersions \cite{2004_Kraynik}, C = 0.3 . 

Eq. \ref{equation_OP} was first proposed in the case of monodisperse foams \cite{2008_Hohler} with $\varphi_c = 0.26$ and $k = 7.3$. It was later shown that it also applies to polydisperse foams and emulsions (polydispersity below 50\%) with another set of constants \cite{2013_Maestro}: $\varphi_c = 0.36$ and $k=3.2$. Note that the parameter $\varphi_c$ is the same in Eq.  \ref{equation_OP} for the osmotic pressure and Eq. \ref{equation_YS_emulsion} for the emulsion yield stress; in both equations, this parameter represent the volume fraction of the continuous phase above which the droplets are spherical \cite{1979_Princen,1983_Princen}.
It can therefore be obtained from the yield stress curve in Fig. \ref{histogram_emulsion} for our emulsions. We have $\varphi_c = 0.36$, in accordance with the result from \cite{2013_Maestro}.

Following the procedure proposed by \textit{Maestro et al.} \cite{2013_Maestro}, we combine Eqs. \ref{equation_integral_dOP} and \ref{equation_OP} to get an analytical relation for $\tilde{z}(\varphi)$ between the reduced height and the volume fraction of the continuous phase: 
\begin{multline}
\tilde{z} =  \dfrac{k}{1-\Phi_P} \left( \sqrt{\varphi_c} - \sqrt{\varphi}  \right)  \left( 3 + (1-\Phi_P) \dfrac{\sqrt{\varphi_c^3}}{\sqrt{\varphi}}\right) \\
+ \dfrac{k}{2 (1-\Phi_P)^{3/2}} (3 - 2 \varphi_c (1-\Phi_P) - \varphi_c^2 (1-\Phi_P)^2) \\
\ln \left(\dfrac{\left(\sqrt{\varphi(1-\Phi_P)}+1 \right) \left(1-\sqrt{\varphi_c(1-\Phi_P)} \right)}{\left(\sqrt{\varphi_c(1-\Phi_P)}+1\right)\left(1-\sqrt{\varphi(1-\Phi_P)}\right)} \right) .
\label{equation_z-phi}
\end{multline}

\begin{figure}[h]
\centering
\includegraphics[width=6.5cm]{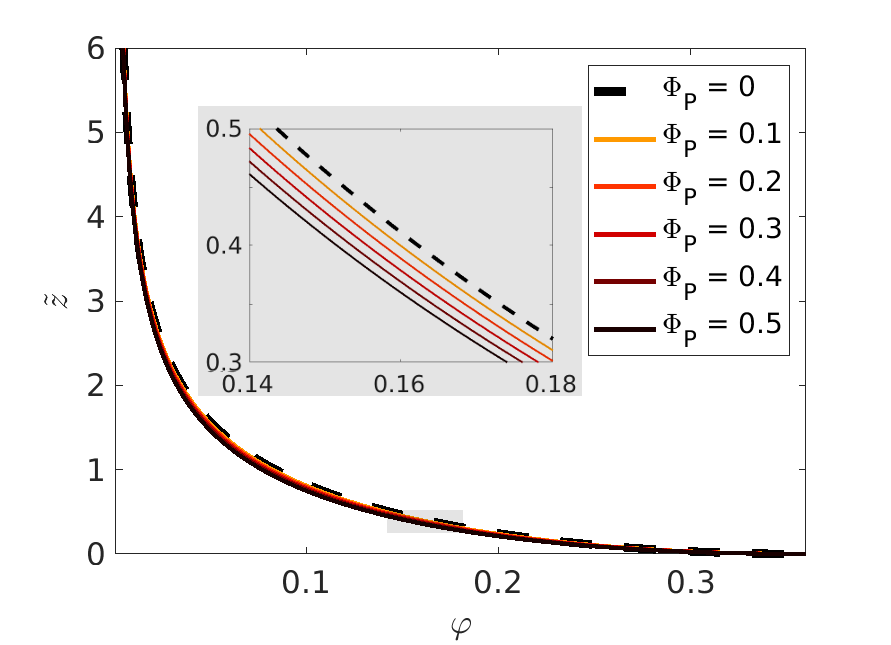}
\caption{The dimensionless liquid fraction profiles for different particle fractions $\Phi_P$, obtained from Eq. \ref{equation_z-phi}. Curve for $\Phi_P = 0$ is the same as in \cite{2013_Maestro}.} 
\label{graph_z-phi_profiles}
\end{figure}

When $\Phi_P = 0$, Eq. \ref{equation_z-phi} is the same formula as obtained by \textit{Maestro et al.} \cite{2013_Maestro}. Eq. \ref{equation_z-phi} is plotted for six values of $\Phi_P$ in Fig. \ref{graph_z-phi_profiles}. The profiles obtained in the presence of particles have a similar shape as the profile without particles. Though, for a given dimensionless height $\tilde{z}$, $\varphi$ decreases when the particle fraction increases. When $\Phi_P = 50\%$, the maximal relative difference with the profile without particles is 20\%.

We note $V_{dest}$ the maximal volume of the continuous phase that a sample can contain without phase separating. $V_{dest}$ is obtained by integrating the total liquid fraction $\Phi$ over the height of the sample, under the assumption that the maximal value of $\varphi$ is $\varphi_{dest}$ at $\tilde{z}_{dest} = \tilde{z}(\varphi_{dest})$. Using Eq. \ref{equation_dOP_2}, we obtain a simple expression for the maximal volume:

\begin{multline}
\tilde{V}_{dest} = \dfrac{R_{32}}{\pi R_{Sample}^2 l_c^2} V_{dest}  = \int_{\tilde{z}_{dest}}^{\tilde{H}+\tilde{z}_{dest}} \Phi d\tilde{z} \\ = \tilde{H} -\left(\tilde{\Pi}(\tilde{H}+\tilde{z}_{dest}) - \tilde{\Pi}(\tilde{z}_{dest})\right),
\label{equation_volume_dest}
\end{multline}
where we assume the sample to be cylindrical with a radius $R_{Sample}$ = 1.3 cm and a height H = 7.5 cm, so that $V_{tot} = 40$ mL.

The theoretical stability diagram is obtained by comparing $V_{dest}$ with the volume of continuous phase in the samples, given by
\begin{equation}
V_{CP} = V_{tot} \varphi_i (1-\Phi_P).
\label{equation_volume_CP}
\end{equation}

\bigbreak

\begin{figure}[h]
\centering
\includegraphics[width=6.5cm]{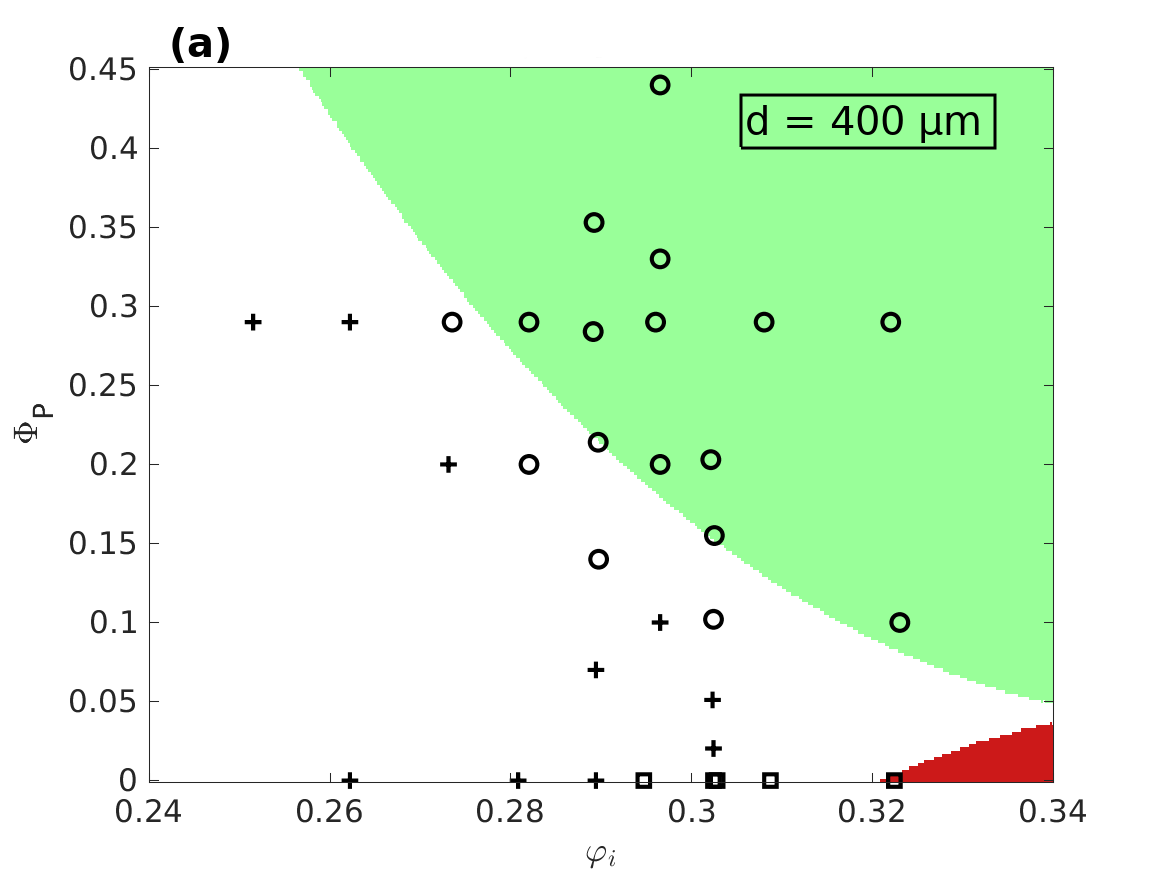}
\includegraphics[width=6.5cm]{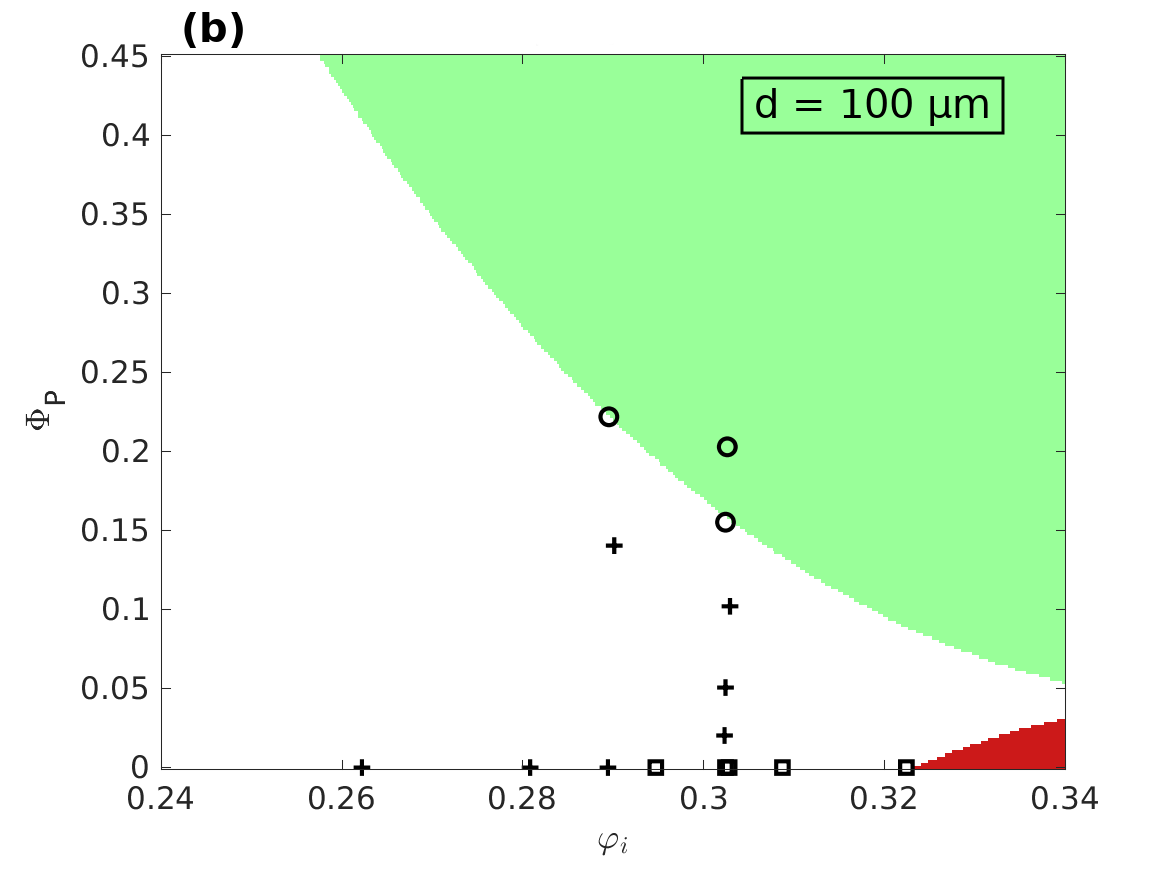}
\caption{Stability of the samples after two months at rest when (a) $d = 400 \mu$m and (b) $d = 100 \mu$m. Black symbols refer to experimental results: crosses indicate no phase separation (configuration (1)), squares shows a layer of liquid is observed at the bottom (configuration (2)), and circles when sedimentation of particles takes place at the top of the sample (configuration (3)). Color of the regions indicate the stability predicted by the static model with $\varphi_{dest}=\varphi_s$ for each particle size: white area for no phase separation, red for particle sedimentation at the bottom of the sample and green for particle sedimentation at the top of the sample.}
\label{graph_results}
\end{figure}

The prediction of phase separation is compared with the experimental results after two months at rest (Fig. \ref{graph_results}). Three different experimental configurations are obtained, which are described below and schematized on the left-hand side of Fig. \ref{schema_method}:

\begin{enumerate}
\item[(1)] \underline{Stable sample}: No phase separation occurs, particles (if present) remain uniformly distributed in the sample and no liquid layer appears.
\item[(2)] \underline{Creaming}: A layer of liquid appears at the bottom of the samples. The color of the liquid indicates that it is the continuous phase.
\item[(3)] \underline{Sedimentation of particles}: A layer of particle-free emulsion appears at the top of the sample. Except for one sample (400 $\mu$m, $\Phi_P = 10.2\%$ and $\varphi_i = 30.2\%$), a layer of continuous phase forms between the particle-free and the particle-laden parts of the sample.
\end{enumerate}

We see in Fig. \ref{graph_results} that the theoretical stability diagram has the same shape as the experimental results. Three distinct final configurations are predicted with the theoretical model for each particle size: no particle sedimentation, particle sedimentation at the top of the sample, and particle sedimentation at the bottom of the samples. For all values of $\varphi_i$, sedimentation of particles at the top of the sample is expected if the particle content is large enough. Indeed, the presence of particles increases the density difference between the dispersed and continuous phase and enhances drainage. The minimal value of $\Phi_P$ above which sedimentation occurs increases when $\varphi_i$ decreases. 

\begin{figure}[ht]
\centering
\includegraphics[width=5.7cm]{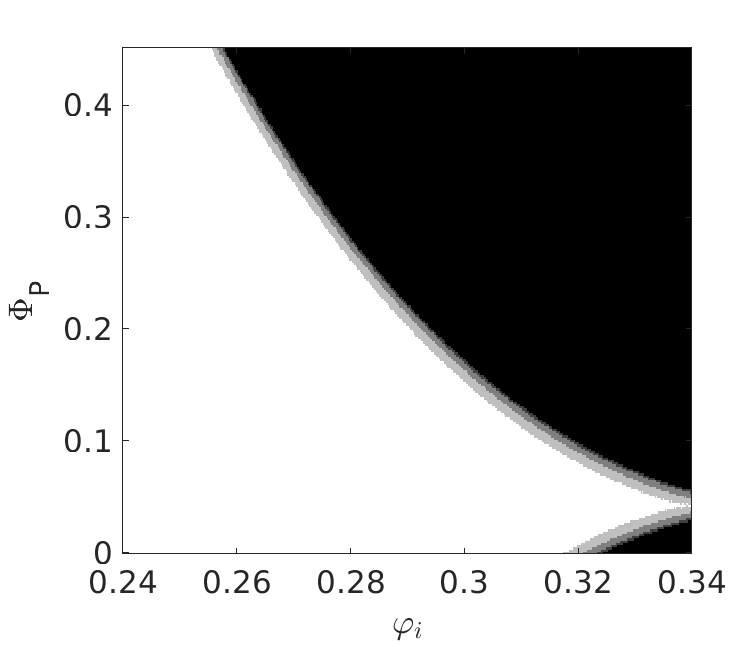}
\includegraphics[width=1.8cm]{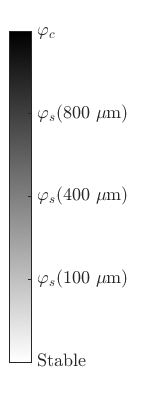}
\caption{Effect on the parameter $\varphi_{dest}$ = [$\varphi_c $, $\varphi_s(100 \mu m)$, $\varphi_s(400 \mu m)$, $\varphi_s(800 \mu m)$] on the theoretical stability diagram. White indicates stable samples (no phase separation) for all four values of $\varphi_{dest}$ and black area, phase separation in all cases. Diagram is obtained form Eqs. \ref{equation_OP}, \ref{equation_z-phi}, \ref{equation_volume_dest} and \ref{equation_volume_CP}, and the values of $\varphi_s$ are calculated with Eqs. \ref{equation_criterion} and \ref{equation_YS_emulsion}. In the case $\varphi_{dest} = \varphi_c = 0.36$, a layer of continuous phase appears at the top or bottom of the unstable samples. The cases $\varphi_s(400 \mu m)$ and $\varphi_s(100 \mu m)$ reflect particle sedimentation. The case $\varphi_s(800 \mu m)$ provides an estimate of particle sedimentation considering sedimentation enhancement due to the large particle number.} 
\label{graph_stability_phiS}
\end{figure}

Although the yield stress threshold predicted by Eq. \ref{equation_criterion} is proportional to the particle diameter, the experimental stability diagrams for 400 $\mu$m and 100 $\mu$m are fairly similar. In addition, experimental results show a remarkable effect: nearly all samples where particle sedimentation occurs also exhibit a layer of continuous phase. To check whether the model can explain these observations, we plot in Fig. \ref{graph_stability_phiS} the theoretical stability diagram obtained for four different values of $\varphi_{dest}$: $\varphi_c$, $\varphi_s(100~\mu m)$, $\varphi_s(400~\mu m)$ and $\varphi_s(800~\mu m)$. In other words, we analyse whether the stability diagram obtained from the model is sensitive to the studied destabilization mechanism and to the particle size. The last $\varphi_{dest}$ value is chosen to take into account possible sedimentation enhancement due to the large number of particles. Indeed, studies have shown that $\tau_{y,s}$ increases when two particles are interacting. Though, in the ``worst" configuration, i.e. two spheres in contact aligned vertically, the value is less than doubled \cite{2009_Jossic,2018_Chaparian}. Our predictions suggest that the stability is nearly insensitive to $\varphi_{dest}$ in this range. One consequence is that the model predicts that particle sedimentation does hardly depend on their size (when the diameter is 400 $\mu$m and below), and that nearly all samples with particle sedimentation, a layer of continuous phase forms at the top or at the bottom. This result is in accordance with our experimental observations.

\begin{figure}[ht]
\centering
\includegraphics[width=6.5cm]{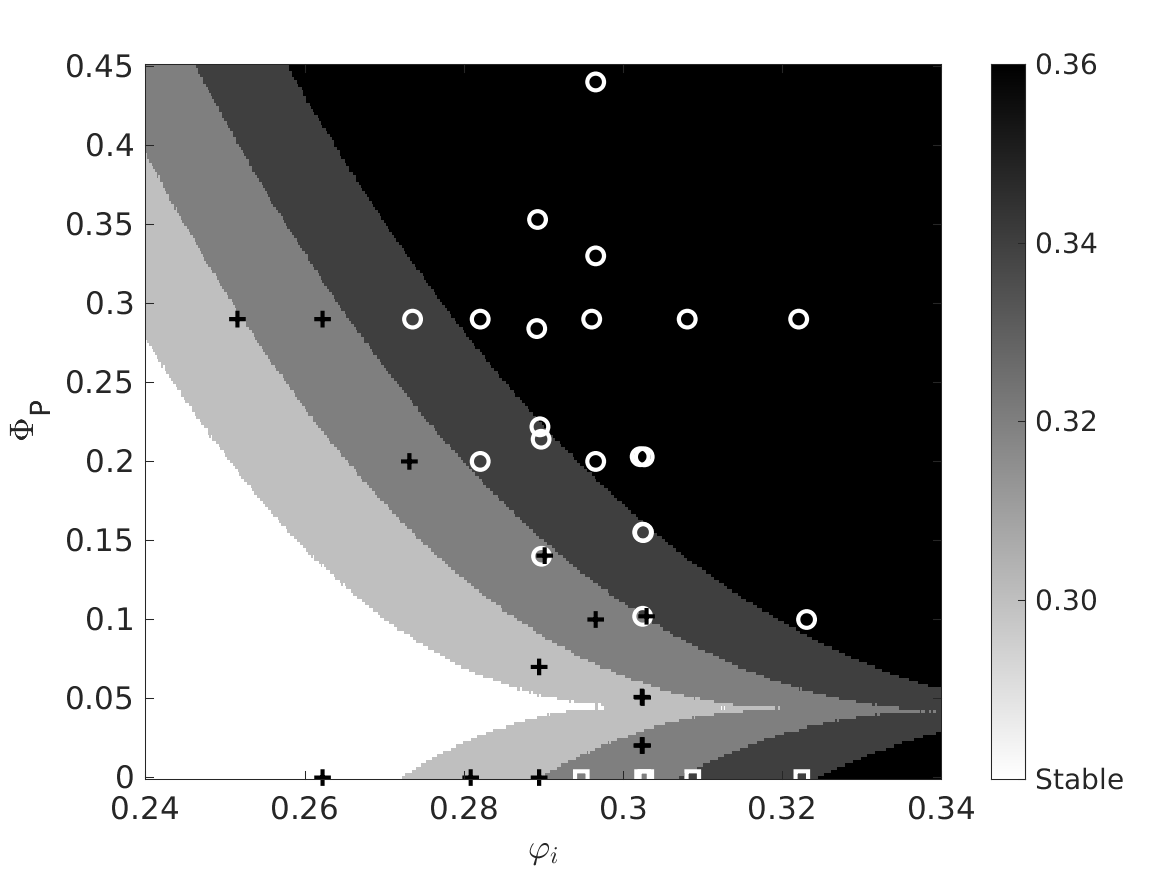}
\caption{Effect of the parameter $\varphi_c = [0.30, 0.32, 0.34, 0.36]$ on the theoretical stability diagram. Diagram is obtained form Eqs. \ref{equation_OP}, \ref{equation_z-phi}, \ref{equation_volume_dest} and \ref{equation_volume_CP}. White indicates always stable samples, and black area, always unstable samples. Here we have taken $\varphi_{dest} = \varphi_c$. Symbols show experimental results, for both particle size: white circle show particle sedimentation, white squares, emulsion creaming, and black crosses, stable samples. } 
\label{graph_stability_phiC}
\end{figure}

\begin{figure*}
\centering
\includegraphics[height = 4.55cm]{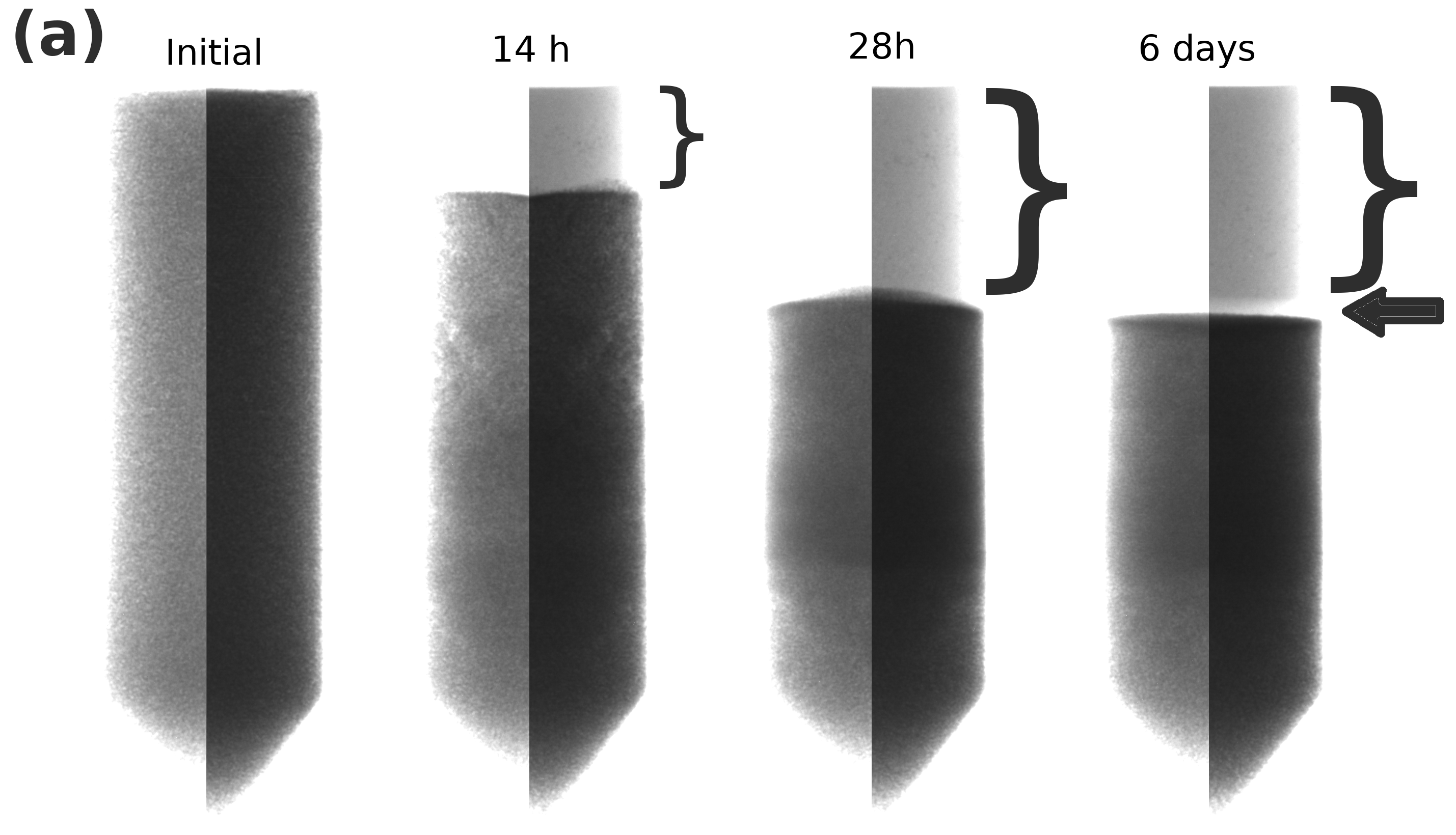}
\includegraphics[height = 4.55cm]{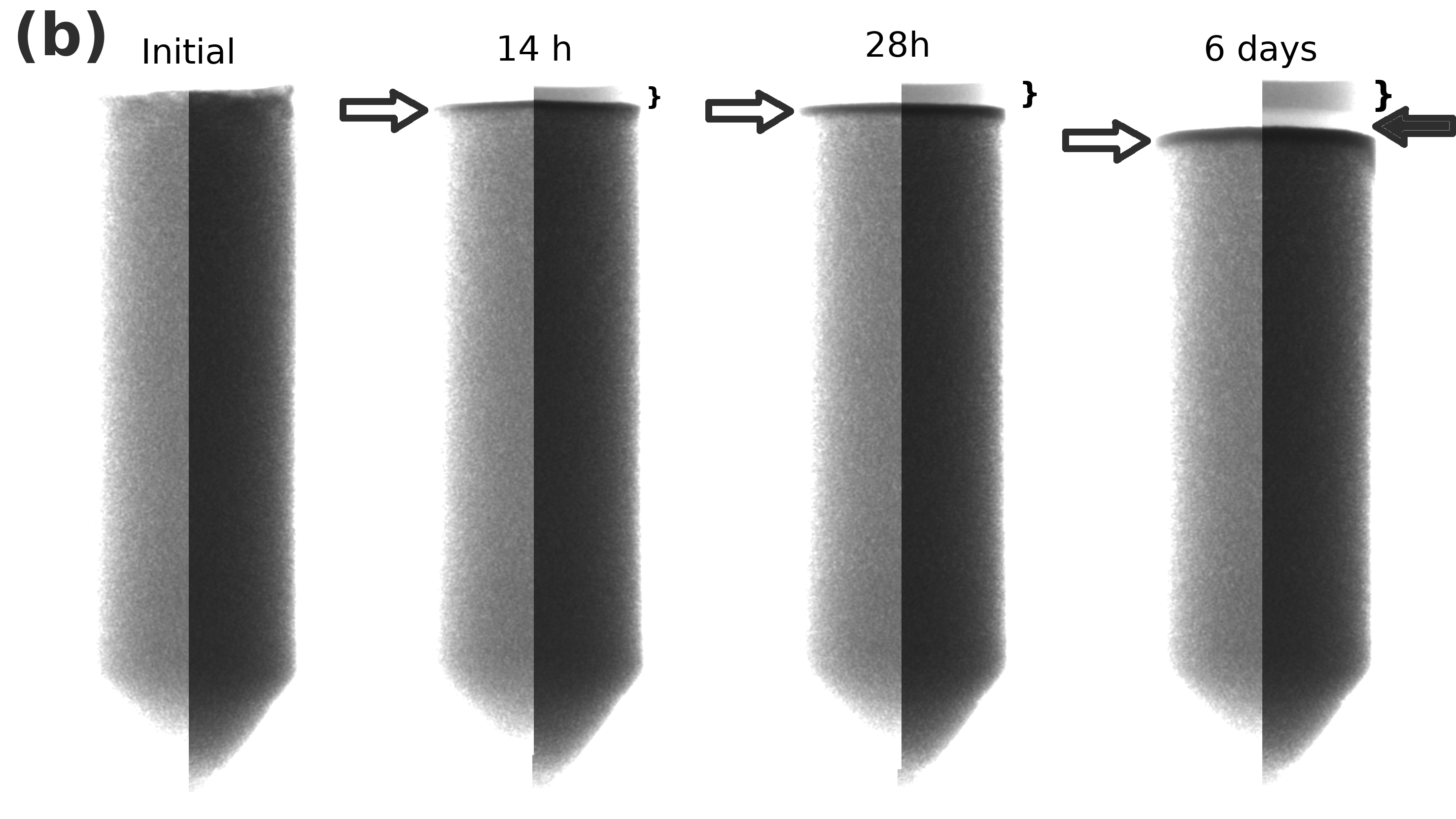}
\caption{Spatio-temporal distribution of particles in unstable samples shown from X-ray measurements. (a) $\Phi_P = 29.1\%$ and $\varphi_i = 32.2\%$. (b) $\Phi_P = 29.0\%$ and $\varphi_i = 30.8\%$. Each X-rays image is composed of two pictures: with high voltage (left half), we can see the distribution of the particles, and especially, on sample (b), the formation of a dense layer indicated by a hollow arrow. With the lower X-ray voltage (right half), we can first see the formation of a light grey layer at the top, with corresponds to a particle-free emulsion layer (semicolumn). Then between the particle-free and the particle-laden emulsion a layer of pure continuous phase appears (black arrow). }
\label{pictures_samples_Xrays}
\end{figure*}

The chosen value for $\varphi_c$, on the other hand, strongly modifies the stability diagram as shown in Fig. \ref{graph_stability_phiC}. In these calculations, we have chosen $\varphi_{dest}=\varphi_c$ for each value of $\varphi_c$. Like in Fig. \ref{graph_stability_phiS} when $\varphi_c = 0.36$, we have checked that for the other values of $\varphi_c$, the stability diagram is fairly insensitive to $\varphi_{dest}$. The stable area is reduced when $\varphi_c$ decreases. Simulations on polydisperse spheres have shown that the packing fraction of the spheres increases with the polydispersity \cite{2009_Farr}. Therefore, we would expect that increasing the polydispersity of the emulsion would enhance particle sedimentation for a given continuous phase content $\varphi_i$. For our experimental samples, it seems that the value $\varphi_c = 0.34$ provides a better fit of our experimental results. This value is lower than reported by \textit{Maestro et al.} \cite{2013_Maestro} (0.36) for foams and emulsions with moderate polydispersity. However, it is still higher than the value reported by \textit{Princen and Kiss} for emulsions \cite{1987_Princen} (0.285). Our yield stress measurement shown in Fig. \ref{graph_YS} give $\varphi_c = 0.36$, and it is surprising that values for the critical volume fraction of the continuous phase are not the same in yield stress measurement and in the drainage-sedimentation tests. This might be related to the time scale of the experiments: yield stress tests take a few minutes, whereas we assess the sample stability at the time scale of weeks. When submitted to stresses slightly below the yield stress, yield stress materials can flow slowly due to creep \cite{2008_Caton}. In the following, we will use the best fit for the stability graph, i.e. $\varphi_c = 0.34$.  

\subsection{Settling dynamics}

\begin{figure*}
\centering
\includegraphics[width=0.23\textwidth]{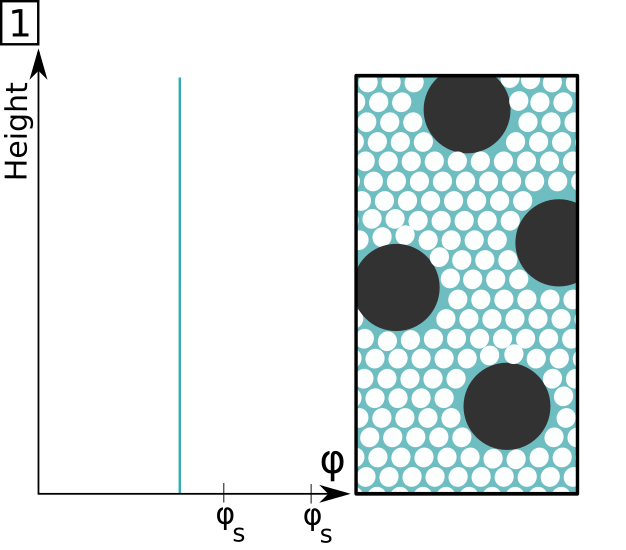}
\includegraphics[width=0.23\textwidth]{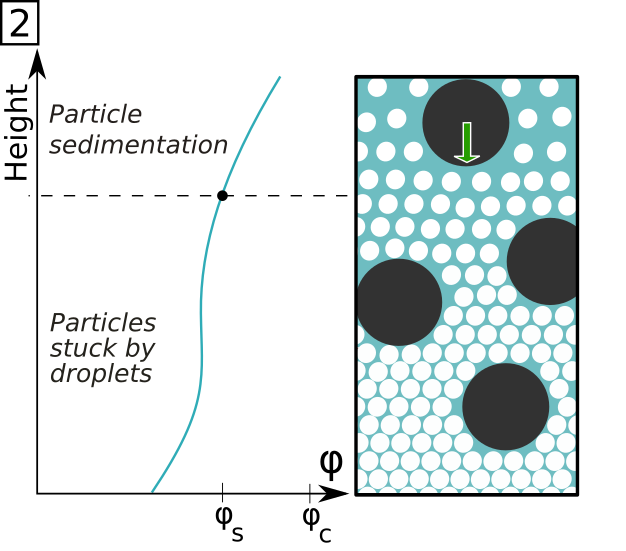}
\includegraphics[width=0.23\textwidth]{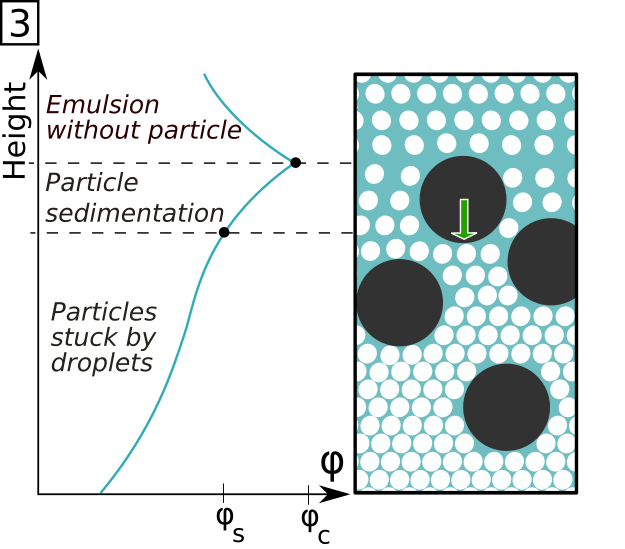}
\includegraphics[width=0.23\textwidth]{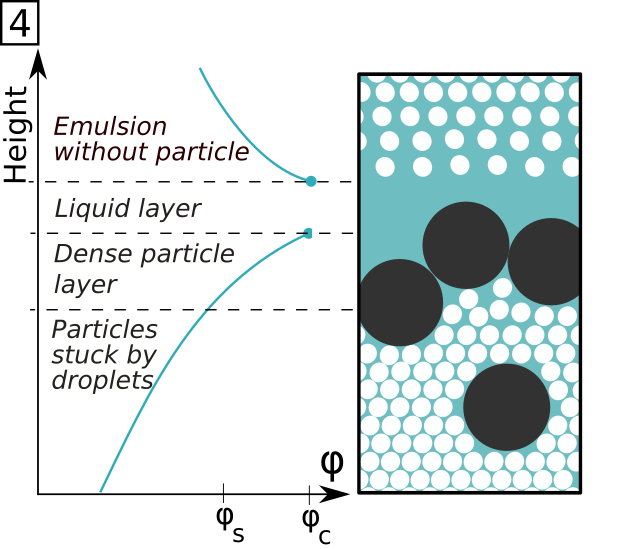}
\caption{Illustration of emulsion drainage, when the dispersed is less dense than the continuous phase, with graphs showing the volume fraction of the continuous phase as a function of height. 1: just after homogenization. 2: drainage leads to an increase of $\varphi$ at the top of the sample. 3: at the top of the sample, $\varphi > \varphi_s$ and particle sediment. 4: sedimentation of particles is stopped by the formation of a dense particle layer. Drainage of both particle-free emulsion in the top part and particle loaded emulsion lead to the formation of a layer of continuous phase between both parts.} 
\label{schema_mechanism}
\end{figure*}

To further understand particle sedimentation in emulsions,  we turn to the dynamic sedimentation process. Fig. \ref{pictures_samples_Xrays} shows the evolution of two samples with time, measured by X-rays. For these samples, like for all samples where particle sedimentation occurs at the top, we first observe a dormant period before the sedimentation starts. Depending on the samples, it lasts between a few hours and a few days. Then, a particle-free layer of emulsion appears at the top of the sample and it increases with time until sedimentation is arrested, mostly due to the formation of a dense layer of particle (see sample (b) in Fig. \ref{pictures_samples_Xrays}). After several days or weeks, a layer of the continuous phase forms below the particle-free layer of the emulsion. A suggestion for a mechanistic description of this destabilization process is illustrated in Fig. \ref{schema_mechanism} when $\rho_d > \rho_{CP}$. Just after the samples are homogeneized, $\varphi$ is uniform. Then, drainage leads to a progressive increase of the volume fraction of the continuous phase at the top of the samples. The dormant period is the time until $\varphi(H) = \varphi_s$. Once $\varphi > \varphi_s$ at the top of the sample, particles sediment and a layer of particle-free emulsion forms. The formation of the layer of continuous phase results from the drainage of both the particle-free top of the sample, where $\rho_d < \rho_{CP}$ and the particle-laden bottom of the sample, where $\rho_d > \rho_{CP}$.

We will focus on the waiting time or dormant period, i.e. the time before the particles start to sediment, denoted by $\tau_w$. In this phase, particles remain uniformly distributed in the sample, so the hypotheses of the model, i.e that particle change the effective density of the dispersed phase without affecting the osmotic pressure, are still valid. Let us now use these hypotheses in the transient state to calculate the time evolution of the $\varphi$ profile. 

When continuous phase flows between the droplets and the particles, the flow velocity $u_{m,CP}$ is given by Darcy's law: 

\begin{equation}
u_{m,CP} = \dfrac{\alpha}{\mu} \left(-\rho_{CP} g - \dfrac{\partial p}{\partial z} \right),
\label{equation_darcy_1}
\end{equation}
where $\alpha$ is the permeability of the porous medium, $\mu = 2.46 \cdot 10^{-3}$ Pa.s is the viscosity of the continuous phase and p is the pressure in the continuous phase.

The permeability $\alpha$ of a porous medium depends primarily on the size of the pores. For an emulsion, the pore size is related to the droplet size $R_{32}$ and the volume fraction of continuous phase $\varphi$. The particles are indeformable and two orders of magnitude larger than the droplets, so we assume that they do not affect the permeability of the samples. We will refer to models for foams, because their permeability has been more widely studied than emulsions. In addition to the pores size, the permeability is related to the morphology of the foam and on the  flow condition at the surface of the droplets, which depends on the chosen surfactant \cite{2004_SaintJalmes}. When large proteins are used as surfactant, the mobility of the interface decreases, the velocity of the continuous phase between the bubbles decreases, and the permeability is reduced. The morphology of the foam or emulsion depends on the volume fraction of the continuous phase $\varphi$: if $\varphi \lesssim 0.01$, continuous phase is contained in long channels, while if $\varphi \gtrsim 0.1$, the length of these channels is small and the continuous phase is present in nodes \cite{2013_Cantat}. In our emulsions, $\varphi \gtrsim 0.1$ and surface mobility is high due to TTAB surfactant. For these reasons, the so-called \textit{node dominated} model given in Eq. \ref{equation_permeability} seems more relevant \cite{2009_Lorenceau}:

\begin{equation}
\alpha(\varphi) = 2.7 \cdot 10^{-3} R_{32}^2 \varphi^{3/2}.
\label{equation_permeability}
\end{equation} 

The motion of the dispersed and the continuous phases in the sample is very slow, so the sample is always at hydrostatic equilibrium during the experiment \cite{2005_Peppin}. The pressure on a representative volume is therefore the hydrostatic pressure $p_H$, which gradient is:

\begin{equation}
\dfrac{\partial p_H}{\partial z} = - [\Phi \rho_{CP} + (1-\Phi) \rho_{d}]g.
\label{equation_hydrostatic_pressure}
\end{equation}

Let us now imagine a horizontal semi-permeable membrane at height z, which can be crossed by the continuous phase but not by the droplets. The total pressure on the sample at height z is the sum of the pressure p on the continuous phase, and the pressure on the membrane. The latter is the osmotic pressure $\Pi$, as defined by Princen \cite{1979_Princen}. Therefore,

\begin{equation}
p = p_{H} - \Pi.
\label{equation_pressure}
\end{equation}

Consequently, Darcy's law can be written as:
\begin{equation}
u_{m,CP} = \dfrac{\alpha (\varphi)}{\mu}\left(-[\rho_{CP}-\rho_d]\left(1-\varphi (1-\Phi_P)\right) g + \dfrac{\partial \Pi}{\partial z}\right).
\label{equation_darcy_2}
\end{equation}

\bigskip

In addition, the continuity equation states 
\begin{equation}
\dfrac{\partial \Phi}{\partial t} = - \dfrac{\partial u_{m,cp}}{\partial z}.
\label{equation_continuity}
\end{equation}

\begin{figure}[t]
\centering
\includegraphics[width=6.5cm]{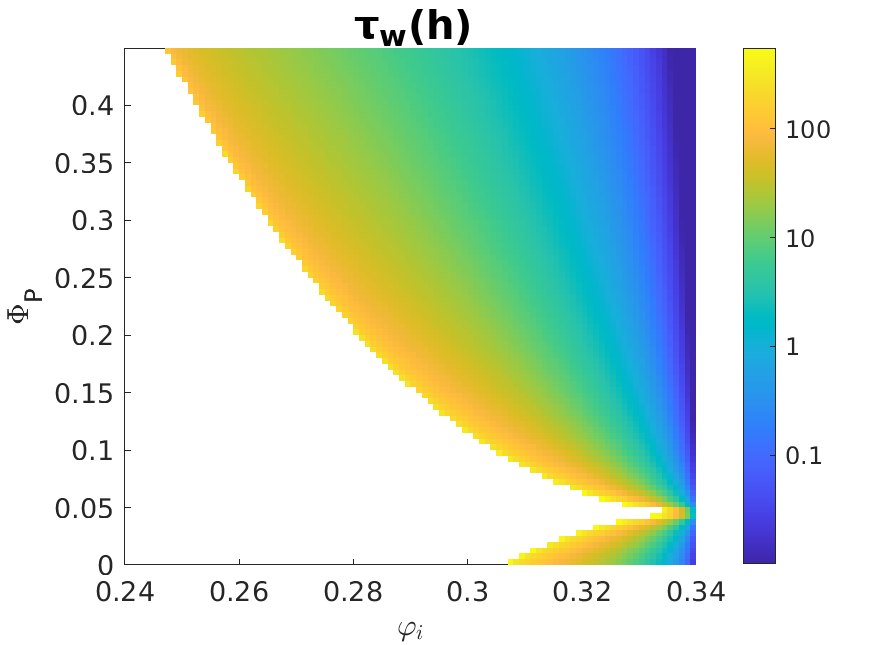}
\caption{Waiting time  $\tau_w$ (in hours) before sedimentation start, predicted from numerical simulations based on Eqs. (\ref{equation_darcy_2}, \ref{equation_continuity}) with $\varphi_s = \varphi_c = 0.34$. White area corresponds to stable samples, i.e. with no sedimentation.}
\label{results_times_diagram}
\end{figure}

We solve explicitly the Eqs. (\ref{equation_darcy_2}, \ref{equation_continuity}) with Matlab, with a time step $\Delta t$ = 0.2 s and a height step $\Delta z = H/1200 = 63~\mu$m, until time t = 1100 hours. In the previous part, we showed that the sample stability prediction is strongly affected by the $\varphi_c$ value, and that $\varphi_c = 0.34$ provides a good fit of our stability graph. The destabilization criterion $\varphi_{dest}$ has a smaller effect. We use therefore in this part $\varphi_c = \varphi_{dest}= 0.34$. The following boundary condition is used at the bottom and the top of the samples: $u_{m,CP}=0$. This gives the evolution of the $\varphi$ profile in time. Sedimentation of particles begins at time $\tau_w$, when at the top or the bottom of the sample, $\varphi = \varphi_{dest}$. We have checked that the value of $\tau_w$ does not depend on the discretization parameters $\Delta t$ and $\Delta z$. $\tau_w$ as a function of the sample composition ($\varphi_i$, $\Phi_P$) is shown in Fig. \ref{results_times_diagram} when $d = 400 \mu$m. As expected, increasing the volume content of the continuous phase $\varphi_i$ at fixed particle content $\Phi_P$ reduces greatly the waiting time $\tau_w$, because of the increase of the permeability $\alpha$. Increasing the particle content increases $\rho_d -\rho_{CP}$ and therefore the hydrostatic pressure $p_H$; thus, it also decreases $\tau_w$.

\begin{figure}[h]
\centering
\includegraphics[width=7cm]{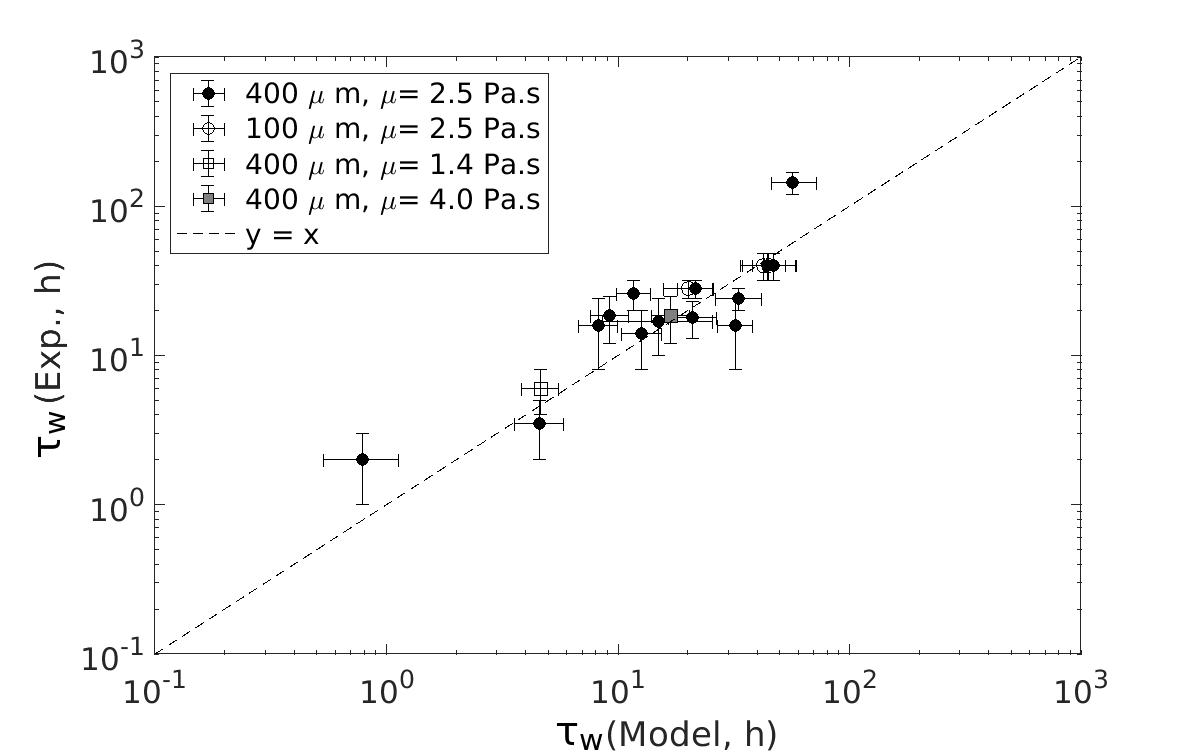}
\caption{Comparison of theoretical and experimental time before sedimentation start for unstable samples. Error bars for experimental results (y-axis) indicate the time of last X-ray picture when sedimentation has not started and the time when particle-free layer is first observed. For model results, we have taken $\varphi_s = \varphi_c = 0.34$, and error bars are the maximum and minimal values obtained with variations $\Delta \varphi_i = 0.2\%$ and $\Delta \Phi_P = 0.5\%$.} 
\label{results_times_modelVSexp}
\end{figure}

In addition to the samples shown in the stability diagram (Fig. \ref{graph_results}), we have prepared two additional samples to check the effect of the viscosity of the continuous phase with $\varphi_i = 29.6\%$ and $\Phi_P = 32.7\%$. These samples are prepared from the same mother emulsion as the other ones, so that the droplet size is not changed, but we have modified the content of glycerol in the additional continuous phase used to dilute the sample. Obtained viscosities are 1.4 Pa.s for 15wt\% glycerol and 4.0 Pa.s for 42wt\% glycerol, in addition to the 2.5 Pa.s for 30wt\% glycerol. The viscosity of the continuous phase does not change the final state of the samples: particle sedimentation occurs in all three samples and the height of the particle-free emulsion is about 4 mm. As expected, $\tau_w$ and the time before the formation of the layer of continuous phase increase with the viscosity of the continuous phase. In Fig. \ref{results_times_modelVSexp}, we compare the experimental length of the dormant period, deduced from the X-ray images, with the predictions from the model. We estimate the sensitivity of the model prediction to the initial volume fraction of continuous phase and particle content by considering small variations $\Delta \varphi_i = 0.2\%$ and $\Delta \Phi_P = 0.5\%$.

We have obtained experimental times for the onset of sedimentation from 1 hour to 100 hours. The model captures well the measured dormant period length for the whole range: the higher the measured values, the higher the model estimations. This confirms that the sedimentation of particles is related to the emulsion drainage.

\section{Conclusions}
We have studied experimentally the sedimentation of particles in emulsions. We observed that the classical criterion for particle sedimentation in yield stress fluids cannot predict particle stability at long time scales (days or weeks). Emulsion drainage leads to a local increase of the volume fraction of the continuous phase, which favors local particle sedimentation.

We propose a simple model to describe the drainage in emulsions containing dense particles. We assume that the particles change the effective density of the dispersed phase, but do not affect the osmotic pressure. Based on these assumptions, we are able to obtain a theoretical stability diagram with the same characteristics as the experimental observations.  First, in contrast to the case of a homogeneous yield stress fluid, we find little effect of the particle size on the sedimentation when the diameter is below 400 $\mu$m. Secondly, for all samples where particle sedimentation occurs, a layer of continuous phase is extracted from the sample. In addition, we observe that adding a moderate quantity of particles can prevent drainage by reducing the density difference between the dispersed and the continuous phase, while all samples with very large particle volume fraction lead to emulsion drainage and particle sedimentation at the top of the sample.

We also model the evolution during the transient regime, i.e. before the profile of continuous phase has reached equilibrium. The waiting time for sedimentation of particle to start is described for the unstable samples. The theoretical times are in good agreement with the experimental measurements.

Materials containing deformable droplets and dense undeformable particles are widely used for industrial application, like drilling fluids in oil industry. Our findings provide a new approach to understanding their stability against gravity.

\section*{Conflicts of interest}
There are no conflicts to declare.

\section*{Acknowledgements}
This work has been carried out as a part of the PIRE project "Multi-scale, Multi-phase Phenomena in Complex Fluids for the Energy Industries", founded by the Research Council of Norway and the National Science Foundation of USA under Award Number 1743794. The authors wish to thank Julie Goyon, Olivier Pitois and Christian Pedersen for fruitful discussions.



\balance

\renewcommand\refname{References}

\bibliography{Bibliography} 
\bibliographystyle{rsc} 

\end{document}